\documentclass[preprint,12pt]{elsarticle}
\usepackage{amssymb}
\usepackage{lscape}
\biboptions{sort&compress}

\begin{document}

\begin{frontmatter}

%% Title, authors and addresses

%% use the tnoteref command within \title for footnotes;
%% use the tnotetext command for the associated footnote;
%% use the fnref command within \author or \address for footnotes;
%% use the fntext command for the associated footnote;
%% use the corref command within \author for corresponding author footnotes;
%% use the cortext command for the associated footnote;
%% use the ead command for the email address,
%% and the form \ead[url] for the home page:
%%
%% \title{Title\tnoteref{label1}}
%% \tnotetext[label1]{}
%% \author{Name\corref{cor1}\fnref{label2}}
%% \ead{email address}
%% \ead[url]{home page}
%% \fntext[label2]{}
%% \cortext[cor1]{}
%% \address{Address\fnref{label3}}
%% \fntext[label3]{}

\title{Cosmogenic production of tritium in dark matter detectors}

%% use optional labels to link authors explicitly to addresses:
%% \author[label1,label2]{<author name>}
%% \address[label1]{<address>}
%% \address[label2]{<address>}

\author{J.~Amar\'e, J.~Castel, S.~Cebri\'an\footnote{Corresponding author (scebrian@unizar.es)}, I.~Coarasa, C.~Cuesta\footnote{Present Address: Centro de Investigaciones Energ\'eticas, Medioambientales y
Tecnol\'ogicas, CIEMAT, 28040 Madrid, Spain}, T.~Dafni,
J.~Gal\'an\footnote{Present address: INPAC and Department of Physics
and Astronomy, Shanghai Jiao Tong University, Shanghai Laboratory
for Particle Physics and Cosmology, 200240 Shanghai, China},
E.~Garc\'ia, J.G.~Garza, F.J.~Iguaz\footnote{Present address: IRFU,
CEA, Universit\'e Paris-Saclay, F-91191 Gif-sur-Yvette, France},
I.G.~Irastorza, G.~Luz\'on, M.~Mart\'inez\footnote{Present Address:
Universit\`a di Roma La Sapienza, Piazzale Aldo Moro 5, 00185 Roma,
Italy}, H.~Mirallas, M.A.~Oliv\'an, Y.~Ortigoza, A.~Ortiz de
Sol\'orzano, J.~Puimed\'on, E.~Ruiz-Ch\'oliz, M.L.~Sarsa,
J.A.~Villar\footnote{Deceased}, P.~Villar}
%\footnote{Corresponding author. Phone number: +34 976761243. Fax number: +34 976761247. E-mail address: scebrian@unizar.es. Postal address: Facultad de Ciencias, Universidad de Zaragoza. Pedro Cerbuna 12, 50009 Zaragoza, Spain.}, H.~G\'{o}mez, G.~Luz\'on, J.~Morales\footnote{Deceased}, A.~Tom\'as, J.~A.~Villar}

\address{Universidad de Zaragoza, Pedro Cerbuna 12, 50009 Zaragoza, Spain\\ Laboratorio Subterr\'aneo de Canfranc, Paseo de los Ayerbe s/n, 22880 Canfranc
Estaci\'on, Huesca, Spain}

\begin{abstract}
The direct detection of dark matter particles requires ultra-low
background conditions at energies below a few tens of keV.
Radioactive isotopes are produced via cosmogenic activation in
detectors and other materials and those isotopes constitute a
background source which has to be under control. In particular,
tritium is specially relevant due to its decay properties (very low
endpoint energy and long half-life) when induced in the detector
medium, and because it can be generated in any material as a
spallation product. Quantification of cosmogenic production of
tritium is not straightforward, neither experimentally nor by
calculations. In this work, a method for the calculation of
production rates at sea level has been developed and applied to some
of the materials typically used as targets in dark matter detectors
(germanium, sodium iodide, argon and neon); it is based on a
selected description of tritium production cross sections over the
entire energy range of cosmic nucleons. Results have been compared
to available data in the literature, either based on other
calculations or from measurements. The obtained tritium production
rates, ranging from a few tens to a few hundreds of nuclei per kg
and per day at sea level, point to a significant contribution to the
background in dark matter experiments, requiring the application of
specific protocols for target material purification, material
storing underground and limiting the time the detector is on surface
during the building process in order to minimize the exposure to the
most dangerous cosmic ray components.
\end{abstract}

\begin{keyword}
%% keywords here, in the form: keyword \sep keyword
Cosmogenic activation \sep Tritium \sep Germanium \sep NaI \sep
Noble gases \sep Rare events
%\sep Radioactive background
%% PACS codes here, in the form: \PACS code \sep code
%\PACS 13.85.Tp \sep 23.40.-s \sep 29.40.Wk \sep 95.35.+d
%% MSC codes here, in the form: \MSC code \sep code
%% or \MSC[2008] code \sep code (2000 is the default)

\end{keyword}

\end{frontmatter}

%%
%% Start line numbering here if you want
%%
% \linenumbers

%% main text
%%\section{}
%%\label{}

\section{Introduction}

%Dark matter projects
Many different efforts are being devoted worldwide to the study of
dark matter which could be pervading the galactic
halo~\cite{baudis2016,klasen2015}. One of the strategies followed is
the direct detection of Weakly Interacting Massive Particles (WIMPs)
proposed to constitute this dark matter, making use of different
kinds of very sensitive radiation detectors~\cite{marrodan2016}. The
expected counting rates from the interaction of WIMPs are extremely
low (of the order of a few events per year and ton of detector or
even below), as it is also the case in the study of other rare
phenomena; therefore, dark matter detectors require ultra-low
background conditions. Operating in deep underground locations,
using active and passive shields and selecting carefully radiopure
materials, reduces very efficiently the background for rare events
experiments~\cite{heusser,formaggio}.

%Activation studies
In this context, long-lived radioactive impurities in the materials
of the set-up induced by the exposure to cosmic rays at sea level
(during fabrication, transport and storage) may be even more
important than residual contamination from primordial nuclides and
become very problematic, depending on the target. For instance, the
poor knowledge of cosmic ray activation in detector materials is
highlighted in~\cite{gondolo} as one of the three main uncertain
nuclear physics aspects of relevance in the direct detection
approach pursued to solve the dark matter problem. In principle,
cosmogenic activation can be kept under control by minimizing
exposure at surface and storing materials underground, avoiding
flights and even using shields against the hadronic component of
cosmic rays during surface detector building or operation. But since
these requirements usually complicate the preparation of experiments
(for example, while crystal growth and detector mounting steps) it
would be desirable to have reliable tools to quantify the real
danger of exposing the different materials to cosmic rays. Direct
measurements (by sensitive screening of exposed materials) and
different calculations of yields have been performed for several
materials in the context of dark matter, double beta decay and
neutrino experiments \cite{cebrian}. Many different studies are
available for germanium
\cite{avignone,miley,norman,barabanov,mei,elliot2010,cebrianap,jian,edelweisscos}
and interesting results have been derived in the last years also for
other detector media like sodium iodide~\cite{cebrianjcap,pettus},
tellurium and tellurium oxide~\cite{te,telozza,wang},
xenon~\cite{schumann,mei2016} or neodymium~\cite{nd1,nd2} as well as
for materials commonly used in the set-ups like copper
\cite{cebrianap,schumann,mei2016,coppers,ivan}, lead
\cite{giuseppe}, stainless steel~\cite{mei2016,coppers}, titanium
\cite{mei2016} and teflon~\cite{mei2016}.

%tritium
Spallation of nuclei by high energy nucleons is a very relevant
process in cosmogenic activation, but other reactions like
fragmentation, induced fission or capture are also important for
some nuclei. Tritium is one of radioactive isotopes that can be
cosmogenically induced in many materials by several production
channels, contributing as a very relevant background in the detector
medium of dark matter experiments due to its decay properties: it is
a pure beta emitter with transition energy of 18.591~keV and a long
half-life of 12.312~y~\cite{DDEP}. Following the shape of the beta
spectrum for the super-allowed transition of $^{3}$H, 57\% of the
emitted electrons are in the range from 1 to 7~keV; these electrons
are typically fully absorbed since most of the dark matter detectors
are large enough. Due to the long half-life of tritium, saturation
activity is difficult to reach; however, even below saturation, as
tritium emissions are concentrated in the energy region where the
dark matter signal is expected, tritium can be important and it is
worth of consideration. Quantification of tritium cosmogenic
production is not easy, neither experimentally since its beta
emissions are hard to disentangle from other background
contributions, nor by calculations, as tritium can be produced by
different reaction channels. Some studies on tritium production in
materials of interest for dark matter experiments can be found in
\cite{avignone,mei,edelweisscos,mei2016,lrt2015anais}; the aim of
this work has been to find a reliable method to quantify the
production rate of tritium in several detector media used in WIMP
direct detection. The production cross sections at the different
energies (the so-called excitation functions) have been selected
over the entire energy range of cosmic nucleons and the calculations
made have been compared with available data.

%materials and experiments
The following materials have been taken into consideration. Natural
isotopic abundances have been assumed unless specifically stated.
\begin{itemize}
\item Germanium is being used as a target for dark matter searches for many years,
either as pure ionization detectors~\cite{igex,cogent,cdex} or in
cryogenic detectors measuring simultaneously ionization and heat
\cite{cdms,edelweiss}. Tritium is highlighted as one of the relevant
background sources in future experiments like SuperCDMS
\cite{supercdms}. A first calculation of tritium production in
germanium was made in~\cite{avignone}, followed by others in
\cite{mei,mei2016}; a part of the unexplained background in the low
energy region of the IGEX detectors could be attributed to tritium
\cite{cebriantaup2003}. Recently, a very detailed quantification of
cosmogenic products including tritium has been made by the EDELWEISS
collaboration~\cite{edelweisscos} and presence of tritium in the
enriched germanium detectors (87\% of $^{76}$Ge and 13\% of
$^{74}$Ge) of the {\sc{Majorana Demonstrator}} focused on the study
of the double beta decay has been reported too~\cite{majorana}.
Therefore, calculations of tritium production in germanium can be
cross-checked with all this available information.

\item NaI(Tl) is the scintillator used in the DAMA/LIBRA experiment,
which has observed an annual modulation effect in the detection rate
with a very high confidence level~\cite{dama}; other experiments and
projects like ANAIS~\cite{anais}, KIMS~\cite{kims} and DM-ICE
\cite{dmice} (now joint in COSINE) or SABRE~\cite{sabre} are
underway in order to confirm this result using the same target.
Production of tritium in this material had not been faced, although
DAMA/LIBRA experiment was able to produce limits to the presence of
this isotope in its detectors~\cite{damalibra}. The commissioning of
the ANAIS (Annual Modulation with NaI(Tl) Scintillators) experiment
with nine detectors having a total NaI(Tl) mass of 112.5~kg is
underway in 2017 at the Canfranc Underground Laboratory (Laboratorio
Subterr\'aneo de Canfranc, LSC), Spain, but the first modules are
fully operative for several years. In the ANAIS detectors, after a
detailed analysis of several cosmogenic products~\cite{cebrianjcap},
the presence of tritium is inferred in order to explain the
differences between the measured background and the background
models~\cite{anaisepjc}. Because tritium is a relevant background in
the region of interest for the annual modulation signal, tritium
production in NaI is further analyzed in this work.

\item In the context of the TREX-DM experiment (TPCs for Rare Event eXperiments for Dark Matter)~\cite{trexdm}, the use of both Ar and Ne
gas is envisaged and therefore the production of tritium in these
two materials has been evaluated too. TREX-DM is an approved
experiment to be installed at the LSC intended to detect low-mass
WIMPs using a Micromegas-based TPC. The most sensitive experiments
in the search for dark matter at present are double phase noble
liquid-gas detectors~\cite{reviewlng}; in the case of xenon target,
tritium and other non-noble radioisotopes are easily suppressed by
specific purification procedures~\cite{luxbkg} and hence have not
been considered in this study.
\end{itemize}

The structure of the paper is as follows: in Sec. \ref{cal} the
process of calculating the production rates of tritium based on
selected excitation functions is presented; results obtained for the
different targets considered are shown and discussed in Sec.
\ref{res}, comparing with available previous estimates; finally,
conclusions are summarized in Sec. \ref{con}.

\section{Calculations} \label{cal}

%%%% projectiles
Cosmogenic activation is strongly dependent on the nucleon flux,
neutron to proton ratio, and energies available. At sea level, for
instance, the flux of neutrons and protons is virtually the same at
energies of a few GeV; however, at lower energies the proton to
neutron ratio decreases significantly because of the absorption of
charged particles in the atmosphere. For example, at 100~MeV this
ratio is about 3\%~\cite{lal}. Consequently, at the Earth's surface,
nuclide production is mainly dominated by neutrons. If materials
were flown at high altitude where cosmic flux is much greater,
energies at play would be larger and activation by protons should
not be neglected. As confirmed by calculations in
\cite{mei2016,wei}, contribution to tritium production by muons is
irrelevant. Then, only neutron activation at sea level has been
taken into account in this work.

%%%% rate and spectrum

\begin{figure}
  \includegraphics[height=.4\textheight]{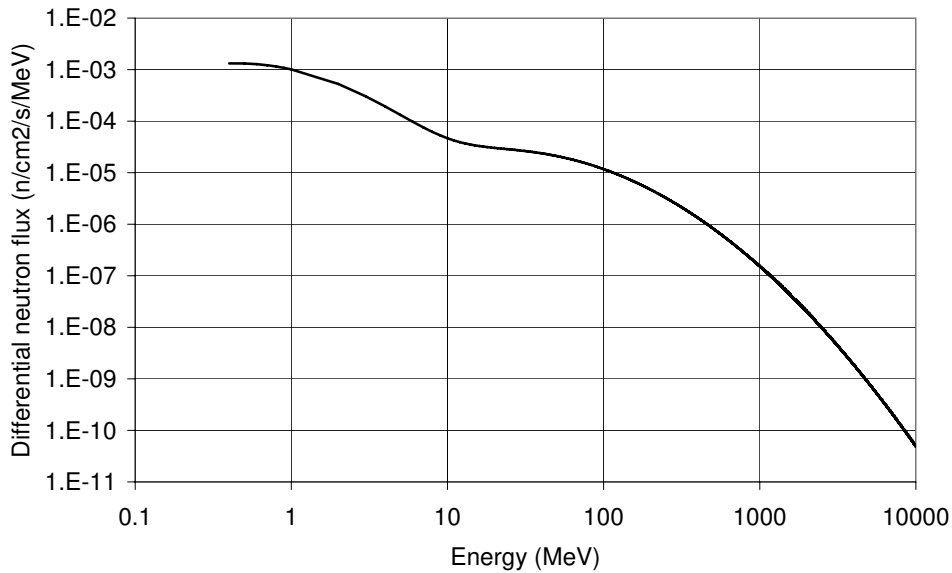}
  \caption{Energy spectrum of cosmic neutrons following the analytic expression
in~\cite{gordon} (for conditions of New York City at sea level)
which has been considered in this work to calculate production rates
of cosmogenic isotopes.}
  \label{nspectrum}
\end{figure}

The flux of cosmic rays ($\phi$) and the production cross-section
($\sigma$) are the two basic ingredients in the calculation of the
production rate $R$ of any isotope by the exposure to a flux
 of cosmic rays, which can be evaluated as:
\begin{equation}
R=N\int\sigma(E)\phi(E)dE \label{eqrate}
\end{equation}
\noindent being $N$ the number of target nuclei and $E$ the particle
energy. For the neutron spectrum at sea level, the parametrization
from~\cite{gordon} has been considered (see figure~\ref{nspectrum}).
It is based on a set of measurements of cosmic neutrons on the
ground across the US, accomplished using Bonner sphere
spectrometers~\cite{bonner}; an analytic expression fitting data for
energies above 0.4~MeV was deduced for reference conditions (New
York City, sea level, mid-level solar modulation)~\cite{gordon}.
Applying this description of cosmic neutrons the integral flux from
10~MeV to 10~GeV is 3.6$\times 10^{-3}$cm$^{-2}$s$^{-1}$.

Concerning the production cross sections, the methodology proposed
in~\cite{cebrian} has been followed: first, collect information from
different sources of data, taken into account both measurements of
production cross sections and calculations using computational
codes; then, select the best description of the excitation function
by nucleons. The following sources have been considered:
\begin{itemize}
\item The EXFOR database (CSISRS in US)~\cite{exfor} provided the
very scarce experimental data of tritium production for Ne, Ar and
Na, all coming from an irradiation experiment with neutrons having
an energy spectrum peaked at 22.5 MeV~\cite{qaim}.
\item  At library TENDL (TALYS-based Evaluated Nuclear Data Library)
\cite{tendl}, cross sections obtained with the TALYS nuclear model
code system were found for neutrons and protons up to 200~MeV for
all the targets. TENDL-2013 was used for neutrons and TENDL-2015 for
protons\footnote{The older version was chosen for neutrons due to
the availability in the ``Tabular production and total cross
sections'' section of the library of the channel of interest for
tritium ($t$) production for the targets of interest ($^{A}Y$):
$^{A}Y(n,x)t$.}.
\item Library HEAD-2009 (High Energy Activation Data)~\cite{head}
merges data for neutrons and protons from 150~MeV to 1~GeV. Here,
data from HEPAD-2008 (High-Energy Proton Activation Data)
sub-library have been considered, obtained using a selection of
models and codes (CEM 03.01, CASCADE/INPE, MCNPX~2.6.0, \dots)
dictated by an extensive comparison with EXFOR data. Only targets
with Z$\geq$12 are considered, and therefore, neither Ne nor Na data
are available in this library.
\item The YIELDX routine (implementing the most updated version
of the popular Silberberg \& Tsao semiempirical formulae
\cite{tsao1,tsao2,tsao3} giving nucleon - nucleus cross sections for
different reactions) allows to compute production cross sections.
Although it can be used for a wide range of targets and products at
energies above 100~MeV, only products as $^{6}$He and heavier can be
evaluated, and therefore, no information on $^{3}$H could be
obtained.
\end{itemize}

\begin{figure}
  \includegraphics[height=.44\textheight]{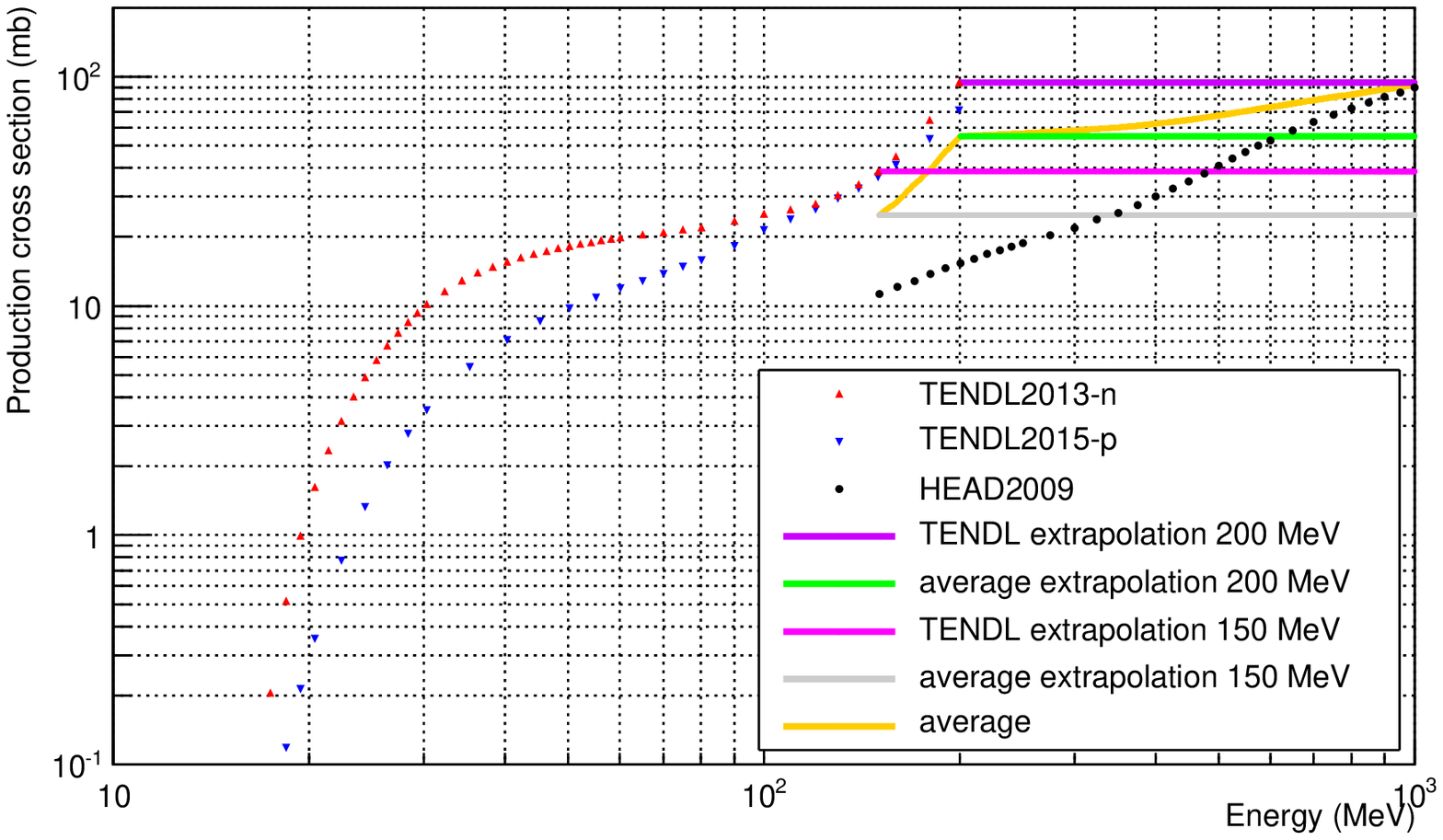}
  \includegraphics[height=.44\textheight]{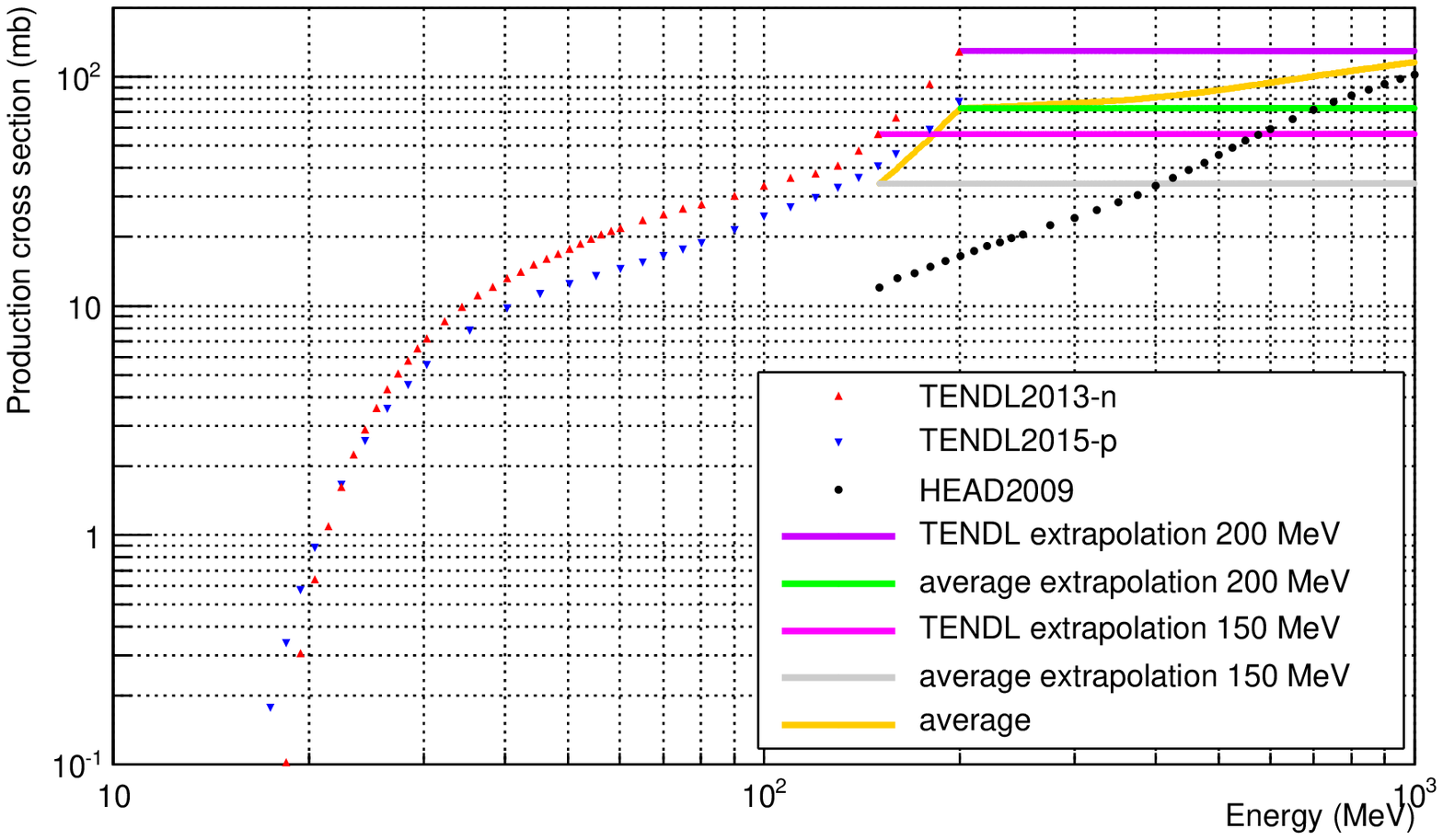}
  \caption{Comparison of excitation functions for the production of $^{3}$H on natural (top) and enriched (bottom) Ge by nucleons taken
  from different sources (TENDL-2013 and HEAD-2009 libraries) together with several extrapolations considered at high energies (see text).
  Above 1~GeV (not shown in the plots), a constant production cross-section from the last available energy has been assumed. An isotopic composition of 87\% for $^{76}$Ge and 13\% for
$^{74}$Ge has been considered for the enriched germanium.}
  \label{efGe}
\end{figure}

\begin{figure}
  \includegraphics[height=.45\textheight]{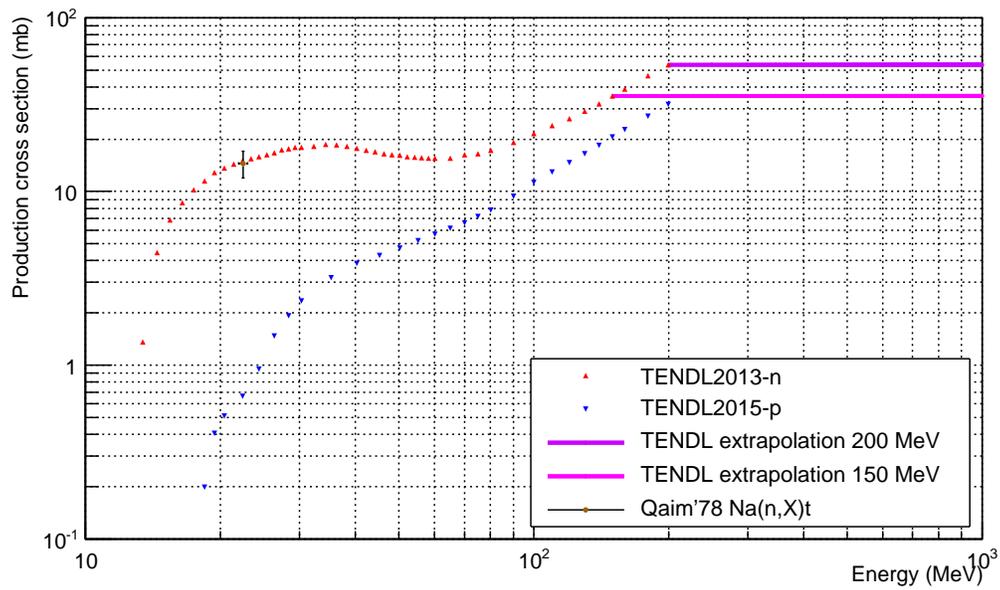}
    \includegraphics[height=.45\textheight]{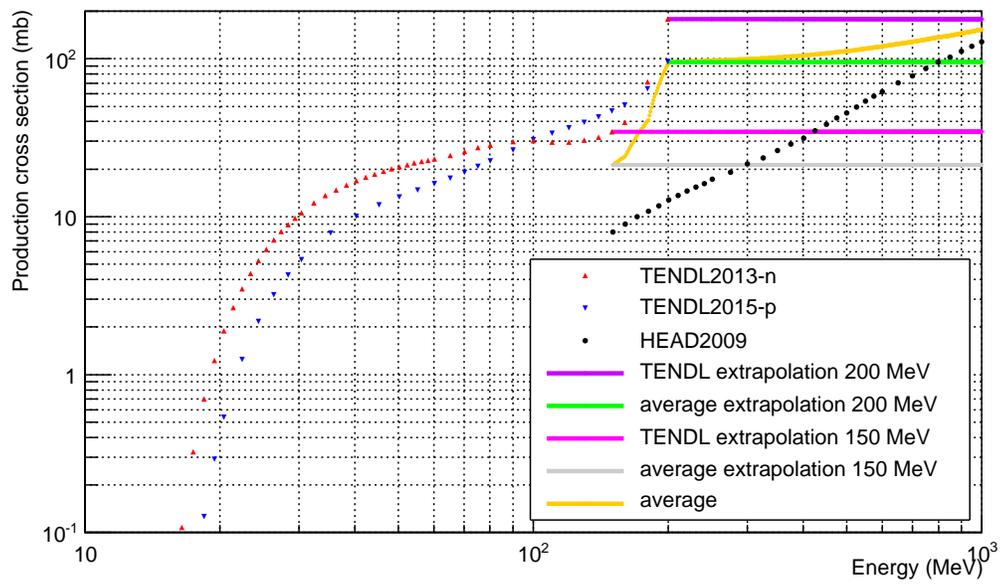}
  \caption{As plots in figure~\ref{efGe}, but for Na (top) and I (bottom) as targets.}
  \label{efNaI}
\end{figure}

\begin{figure}
  \includegraphics[height=.45\textheight]{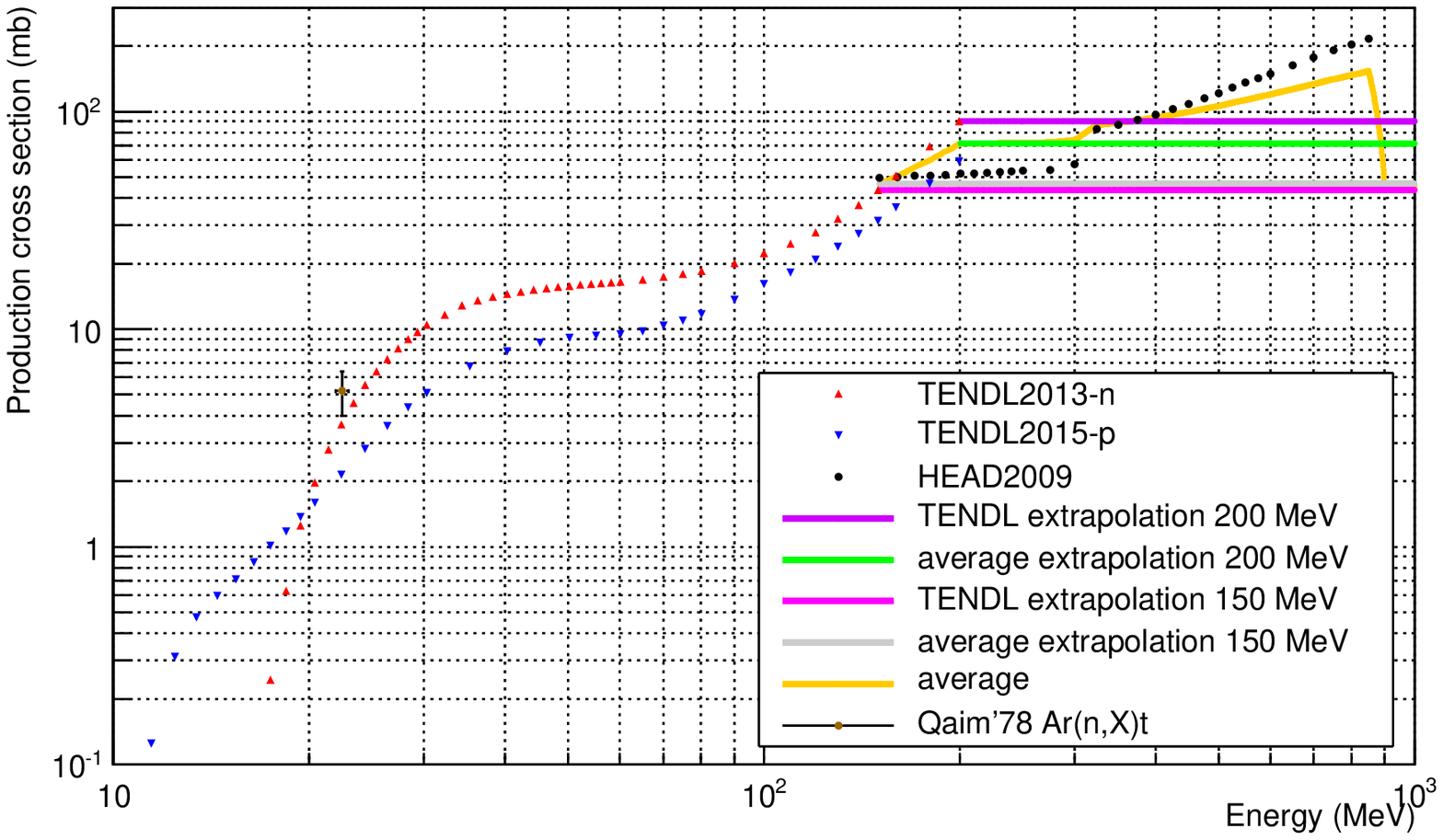}
  \caption{As plots in figure~\ref{efGe}, but for Ar as target.}
  \label{efAr}
\end{figure}

\begin{figure}
  \includegraphics[height=.45\textheight]{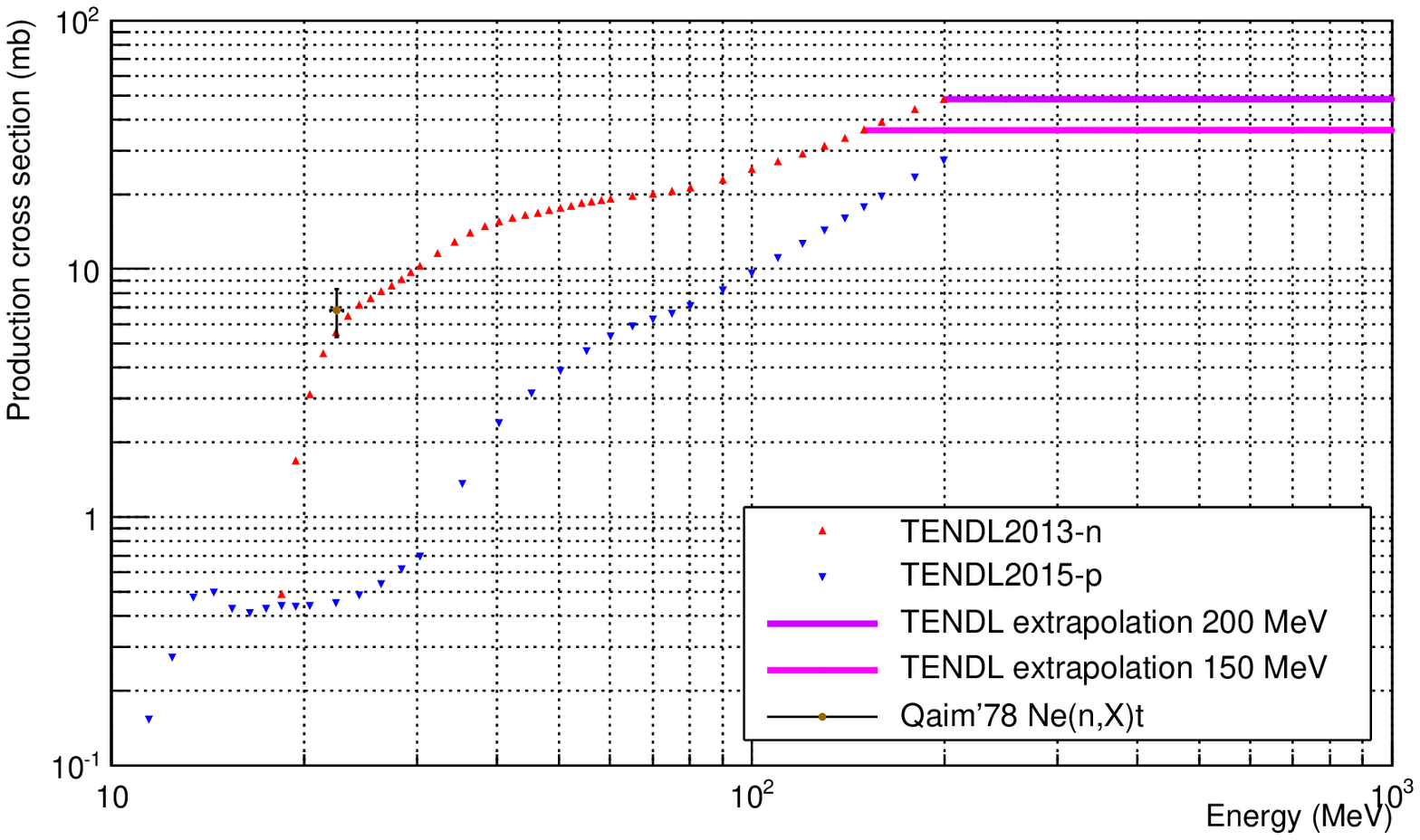}
  \caption{As plots in figure~\ref{efGe}, but for Ne as target.}
  \label{efNe}
\end{figure}

Figures~\ref{efGe}-\ref{efNe} show all the available information on
the excitation function by nucleons collected for the different
analyzed targets, together with some extrapolations in the high
energy range which will be described in the next section.
Independent results for neutrons and protons are presented whenever
possible, to compare the corresponding cross sections, even if
proton activation has not been evaluated here. Production rates have
been computed following equation~\ref{eqrate}, by convoluting a
particular excitation function with the described energy spectrum of
cosmic neutrons at sea level; the energy step in the calculations
has been 0.1~keV for TENDL data integration, 1~keV for HEAD-2009
data up to 1~GeV and 100~keV at higher energies.

\section{Results and discussion}
\label{res}

In order to choose the best description of excitation functions, it
would be desirable to evaluate deviation factors between calculated
and experimental cross sections; for the tritium production
considered here, unfortunately, there is only one experimental point
for Ar, Na and Ne (see section~\ref{cal}). For the lowest energies,
the TENDL data for neutrons up to 200~MeV reproduce well the
measurements in the three cases. At higher energies, as there is no
experimental data to validate the cross section selection, and the
trend of the cross-section values suggests that the contribution
cannot be neglected, different assumptions have been considered.
Then, production rates have been derived for different conditions in
the calculation of the integral in equation~\ref{eqrate} and are
shown in table~\ref{ratestritium}. For the higher energies, above
150/200~MeV, the evaluated options are the following:
\begin{enumerate}
\item HEAD-2009 cross sections have been used, if available. It is assumed that in this
high energy range neutron and proton cross sections are comparable.
\item The available highest energy cross-section from TENDL has been considered as a constant value at
higher energies.
\item The average between the available highest energy cross-section from
TENDL and the corresponding results from HEAD-2009 has been
considered also as a constant value at higher energies.
\item The average at each energy between TENDL data (and its
extrapolation) and HEAD-2009 cross sections has been taken into
consideration too.
\end{enumerate}
In table~\ref{ratestritium} and figures~\ref{efGe}-\ref{efNe}, for
brevity sake, option (1) is referred as HEAD-2009, option (2) as
TENDL extrapolation, option (3) as average extrapolation and option
(4) as average.

No data have been found above 1~GeV. Although the cosmic neutron
spectrum decreases quickly with energy (see figure~\ref{nspectrum}),
cross sections seem to be still increasing with energy (see
figures~\ref{efGe}-\ref{efNe}) and then calculations of production
rates have been extended up to 10~GeV; since is difficult to predict
the particular shape of the excitation functions in this energy
range, a constant production cross-section from the last available
energy has been assumed, giving actually a lower limit to the
contribution to the production rates from neutrons of those
energies. As it will be shown afterwards, that contribution is
estimated to be low enough to make this approximation not critical.

\begin{landscape}
\begin{table}
\begin{tabular}{l|cccccc}
\hline & $^{nat}$Ge & $^{enr}$Ge & Na & I & Ar & Ne \\
\hline

LE TENDL & 28.2/40.7 & 31.6/48.8 & 14.3/18.7 & 15.4/22.9 & 47.7/71.8
& 101.5/133.1 \\ \hline

HE (1) &   26.6/23.9 & 28.5/25.6 & & 14.6/13.6 & 111.5/92.3 & \\

HE (2) &  32.8/60.5 & 45.8/79.5 & 14.6/16.6 & 8.7/55.2 & 67.4/105.4 & 110.6/111.5\\

HE (3) & 21.2/35.2& 27.8/44.8 & & 14.2/29.6 & 71.9/82.9 &  \\

HE (4) & 49.8/42.2& 62.6/52.6 & & 38.6/34.4 & 120.5/98.9 &  \\
\hline

total (1) & 54.8/64.6 & 60.1/74.5 & & 30.0/36.5 & 159.3/164.1 &\\

total (2) & 61.0/101.3& 77.4/128.3 & 28.8/35.2& 29.6/78.1 & 115.1/177.2 & 212.1/244.6\\

total (3) & 49.4/75.9 & 59.4/93.6 & & 24.1/52.5 & 119.7/154.7 &\\

total (4) & 78.0/83.0 & 94.2/101.4 & & 54.0/57.3 & 168.2/170.7 &  \\
\hline

estimated rate & 75$\pm$26 & 94$\pm$34 & 32.0$\pm$3.2 & 51$\pm$27 &
146$\pm$31 & 228$\pm$16 \\ \hline
\end{tabular} \caption{Production rates (in kg$^{-1}$d$^{-1}$) of
$^{3}$H at sea level calculated for the considered targets using
different excitation functions. Contributions from low energy (LE,
from the TENDL-2013 library~\cite{tendl}) and high energies (HE, see
text, (1) is for HEAD-2009, (2) for TENDL extrapolation, (3) for
average extrapolation and (4) for average) have been evaluated
cutting at both 150/200~MeV and summed to derive the total
production rates. The final estimated rates are given by the ranges
defined between the maximum and minimum obtained rates. Results for
Na and I are expressed per mass unit of the NaI detector and then
the production rate for NaI is the sum of Na and I values,
corresponding to (83$\pm$27)~kg$^{-1}$d$^{-1}$. }
\label{ratestritium}
\end{table}
\end{landscape}

In the range between 150 and 200~MeV there are (at least for some
targets) cross sections from both TENDL and HEAD-2009. But
unfortunately, there is an important mismatch between excitation
functions from the two different libraries used at lower and higher
energies, especially for germanium and iodine (see
figures~\ref{efGe} and \ref{efNaI}). The lack of experimental
results on cross sections here makes difficult to choose one
description. This introduces an important uncertainty in the
estimate of the production rate; to take into account and to
quantify this effect, the whole process of calculating the
production rate by summing the contribution from low energy (using
TENDL) and that from high energy (for each of the described four
options) has been made twice, considering the cut between low and
high energy either at 150 or at 200~MeV. Table~\ref{ratestritium}
summarizes all the obtained production rates in the eight different
conditions. The maximum and minimum rates define an interval, whose
central value and half width have been considered as the final
results and their uncertainties for the evaluation of the production
rates of tritium in the different targets. For NaI, rates derived
for Na and I have been properly summed.

%the sum between the
%contribution from low energy using TENDL and that from high energy
%considering one of the described four options has been repeated
%considering the cut both at 150 and at 200~MeV.
It is worth noting that the actual production rates must be higher
than the values reported in table~\ref{ratestritium} since they
correspond only to neutron activation. Proton activation, being much
smaller, is not completely negligible. As pointed out in
section~\ref{cal}, neutron and proton fluxes are similar on the
Earth surface for energies above 1~GeV. The relative contribution in
the total production rate of neutrons from this very high energy
range is for instance between 4 and 7\% for all the analyzed targets
when considering cross sections from TENDL extrapolation (total (2)
in table~\ref{ratestritium}) and 10\% and 13\% for germanium and
iodine respectively when assuming the HEAD-2009 cross sections
(total (1) in table~\ref{ratestritium}). Therefore, at least a
similar relative contribution would be expected from protons. These
percentages agree with the results in~\cite{barabanov,wei} where
proton activation in germanium was specifically evaluated for some
isotopes and with the typical contribution from protons to isotope
production quoted as $\sim$10\% in~\cite{lal}.

Table~\ref{ratestritiumlit} compares our estimated rates with the
available information on tritium production rates from the
literature. Results from~\cite{mei} were estimated generating the
excitation functions with TALYS 1.0 code; TALYS ~\cite{talys} is a
software package for the simulation of nuclear reactions that can be
used in the 1~keV to 200~MeV incident energy range. Results from
\cite{mei2016} are based on GEANT4 simulation or ACTIVIA
calculations~\cite{activia} considering neutrons from thermal
energies to 100~GeV; for GEANT4 simulations the set of
electromagnetic and hadronic physics processes included in the
Shielding modular physics list were taken into account while in
ACTIVIA cross sections are obtained from data tables and
semiempirical formulae~\cite{tsao1,tsao2,tsao3}. It is worth noting
that estimates in \cite{mei,mei2016} used the same cosmic neutron
spectrum considered in this work from \cite{gordon} (which required
a modification of the ACTIVIA code when using this package), so
discrepancies between estimates cannot be assigned to the cosmic ray
flux.
%The only experimental production rate deduced for germanium in~\cite{edelweisscos} is also included.

\begin{table}
\begin{tabular}{l|ccccc}
\hline & $^{nat}$Ge  & $^{enr}$Ge & NaI & Ar & Ne \\ \hline
This work & 75$\pm$26  & 94$\pm$34 & 83$\pm$27  & 146$\pm$31 & 228$\pm$16  \\
Measurements &  82$\pm$21~\cite{edelweisscos} &  140$\pm$10~\cite{majoranalrt} & & & \\
& 76$\pm$6 \cite{cdmslite}& & & & \\
TALYS~\cite{mei} & 27.7 & 24.0 & 31.1 & 44.4 & \\
GEANT4~\cite{mei2016} & 48.3 & &  42.9 & 84.9 & \\
GEANT4~\cite{wei} & & 51.3 & & & \\
ACTIVIA~\cite{mei2016} & 52.4 & & 36.2 &82.9 & \\
ACTIVIA~\cite{wei} & & 47.4 & & & \\
ACTIVIA & 46/43.5~\cite{edelweisscos} &  &   26~\cite{pettus} & & \\
COSMO~\cite{cebriantaup2003} & &  70 & & & \\
Ref.~\cite{avignone} &  178/210 & 113/140 & & & \\ \hline
\end{tabular}
\caption{Comparison of the production rates (in kg$^{-1}$d$^{-1}$)
of $^{3}$H at sea level evaluated in this work with available
information from the literature for the considered targets. The two
values from~\cite{avignone} were derived using two different neutron
spectra (normalized to sea level at 45$^{\circ}$ north latitude) and
the two values from ACTIVIA in~\cite{edelweisscos} correspond to
using just semiempirical cross sections or data from MENDL-2P too.
%highest: Hess
} \label{ratestritiumlit}
\end{table}

A discussion of all the results is given in the following for each
one of the considered targets.

\subsection{Germanium}

The first experimental estimate of tritium production rate in
natural germanium at sea level has been presented by the EDELWEISS
collaboration~\cite{edelweisscos}, following a detailed analysis of
a long measurement with many germanium detectors; their exposure
history above ground during different steps of production and
shipment is well-known and has been considered. Another experimental
evaluation of the tritium production rate based on CDMSlite data
\cite{cdmslite} has been presented too, fully compatible with that
of EDELWEISS. An estimate of the rate was made using ACTIVIA code
and the spectrum from \cite{gordon} by EDELWEISS
~\cite{edelweisscos}. Several calculations had been made before: a
rough calculation was attempted in \cite{avignone} using two
different neutron spectra; more recent calculations
from~\cite{mei,mei2016} using different approaches are shown in
table~\ref{ratestritiumlit} together with that presented in this
work. All these new calculations summarized in
table~\ref{ratestritiumlit} give lower values than the measured
rate; in particular, the smallest value from~\cite{mei} can be
understood because using TALYS cross-sections only contributions
from the lowest energy neutrons are considered. The range derived in
this work for the production rate and presented in
table~\ref{ratestritium} is well compatible with the measured rates
by EDELWEISS and CDMSlite.

For enriched germanium, as used in double beta decay experiments,
there are calculations in~\cite{avignone} and using the COSMO code
\cite{cosmo} in~\cite{cebriantaup2003}. In~\cite{wei}, the
methodology applied in~\cite{mei2016} to evaluate production rates
using GEANT4 and ACTIVIA has been used not only for
natural\footnote{Results on tritium production rates in natural
germanium for neutron activation in~\cite{wei} are virtually the
same as those presented in~\cite{mei2016} and for this reason they
are not reported again in table~\ref{ratestritiumlit}.} but also to
enriched germanium. A first estimate of the production rate from the
data of the {\sc{Majorana Demonstrator}}, whose enriched detectors
have a very well-known exposure history, has been recently presented
\cite{majoranalrt}; the fitting model to derive the abundance of
cosmogenic products is comprised of a calculated tritium beta-decay
spectrum, flat background, and multiple X-ray peaks. In this work,
the tritium production rate has been evaluated not only for natural
but also for enriched germanium. All the results for the enriched
material are compared in table~\ref{ratestritiumlit}. According to
measured rates and results obtained in this work, production is
higher than in natural germanium; this is due to the fact that cross
sections increase with the mass number of the germanium isotope in
all the energy range above $\sim$50~MeV, according to TENDL-2013 and
HEAD-2009 data.

\subsection{Sodium Iodide}

Only calculations of tritium production rate in NaI are available;
results from~\cite{mei,mei2016} using different approaches are shown
in table~\ref{ratestritiumlit} together with the estimate from this
work. An estimate using ACTIVIA and the spectrum from \cite{gordon}
from 1~MeV to 10~GeV \cite{pettus} is also presented. As for
germanium, the TALYS estimate gives a low rate which is in
reasonable agreement with the contribution from just low energy
neutrons according to TENDL cross sections derived in this work (see
table~\ref{ratestritium}). As shown in figure~\ref{efNaI}, bottom,
in the medium energy range there is a significant difference between
HEAD-2009 cross sections and the value extrapolated from TENDL,
which is much higher; this fact is responsible of the large
dispersion in the estimated production rates from different
excitation functions for I. ACTIVIA code is based on semiempirical
formulae derived from proton cross sections, as those compiled in
HEAD-2009 library; this could be one reason why ACTIVIA estimates
give lower production rates than those obtained in this work.

\begin{figure}
  \includegraphics[height=.45\textheight]{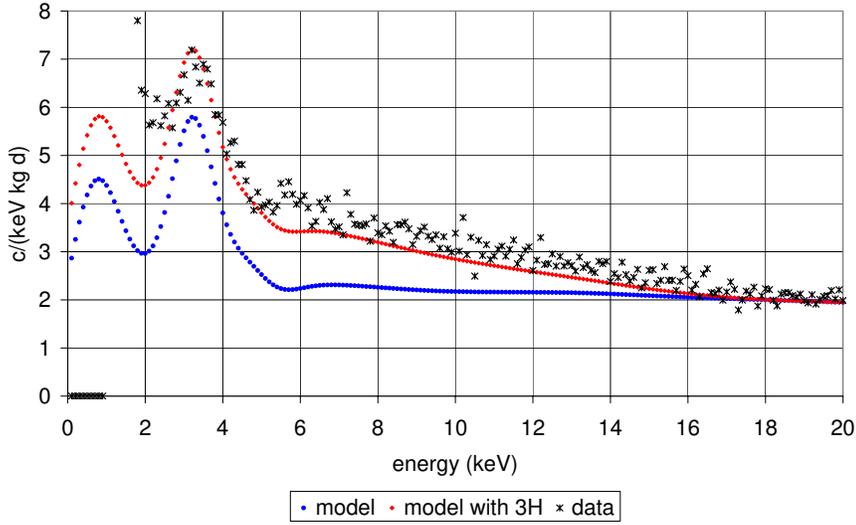}
  \includegraphics[height=.45\textheight]{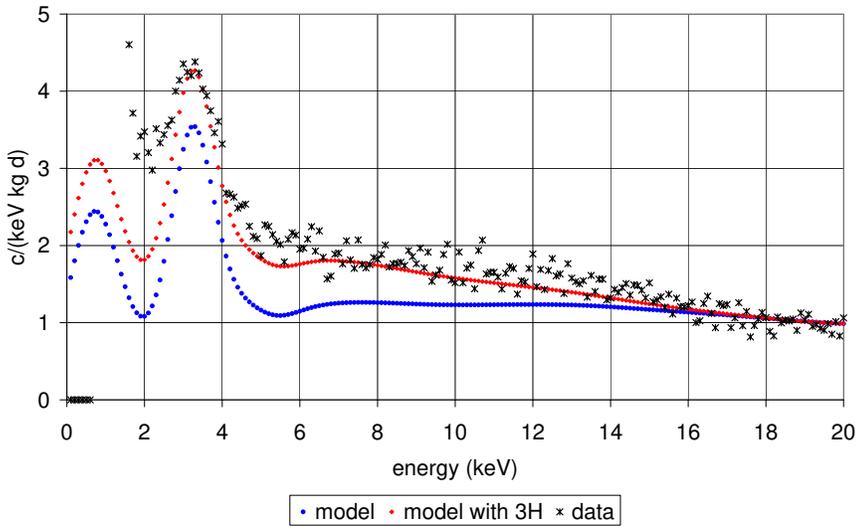}
  \caption{The very low energy region of the energy spectra measured for D0 (top) and D2 (bottom) ANAIS detectors compared with the corresponding
  simulated models~\cite{anaisepjc} including all the quantified intrinsic and cosmogenic activities in the detectors and main components
  of the set-up (blue) and adding also tritium in the NaI crystal (red). A $^{3}$H activity of 0.20~mBq/kg is considered for D0 and 0.09~mBq/kg for D2.}
  \label{anaisspc}
\end{figure}

%ANAIS results
The first ANAIS modules have been taking data at the LSC for several
years in order to study the detector response and background. Using
data taken with the first two 12.5~kg NaI(Tl) detectors produced by
the Alpha Spectra company for the ANAIS experiment, production rates
of several I and Te isotopes and of $^{22}$Na could be derived
\cite{cebrianjcap}, thanks to the very fast start of data taking
after moving the detectors underground. Although a direct
identification of a tritium content in the crystals has not been
possible, the construction of a detailed background model of these
modules and those produced afterwards (based on a Geant4 simulation
of quantified background components) points to the need of an
additional background source contributing only in the very low
energy region, which could be tritium (see details at
\cite{anaisepjc},\cite{tesispatricia}). The simulated spectra
including all well-known contributions agree reasonably with the
ones measured, except for the very low energy region; as shown in
figure~\ref{anaisspc}, the inclusion of a certain activity of
$^{3}$H homogeneously distributed in the NaI crystal provides a very
good agreement also below 20~keV. Figure~\ref{anaisspc} compares
data and background models for two detectors, named D0 and D2, with
different production history. D0 arrived at LSC in December 2012 and
D2 in March 2015, and both have been taking data there since then in
several set-ups. The shown data correspond to 59.9~days of
measurement in September and October 2016. No fitting has been
attempted, but the required $^{3}$H initial activities (that is, at
the moment of going underground) to reproduce the data would be
around 0.20~mBq/kg for D0 and 0.09~mBq/kg for D2; as explained
below, the different values for the two detectors can be understood
due to a different time of exposure to cosmic rays for each
detector. The value estimated for D2 agrees with the upper limit set
on tritium activity for DAMA/LIBRA crystals~\cite{damalibra}.
Preliminary background models developed for the other ANAIS
detectors already operated in Canfranc point to a tritium content
similar to that assumed for D2.
%The latter value is just the upper limit set for DAMA/LIBRA crystals~\cite{damalibra}.

Since the exposure history of the NaI material used to produce the
crystals (following different procedures for purification and
crystal growth) is not precisely known, no attempt of deriving
tritium production rates from these estimated activities in ANAIS
crystals has been made. But some cross-check of the results is
possible to confirm the plausibility of the tritium hypothesis;
assuming that the whole crystal activation took place in Alpha
Spectra facilities at Grand Junction, Colorado (where the cosmic
neutron flux is estimated to be a factor $f=3.6$ times higher than
at sea level~\cite{cebrianjcap}) the required exposure time
$t_{exp}$ to produce an activity $A$ of an isotope with decay
constant $\lambda$ for a production rate $R$ at sea level can be
deduced using
\begin{equation}
A = f R [1-\exp(-\lambda t_{exp})]        %\exp(-\lambda t_{cool})
\label{eqac}
\end{equation}
%a cooling time $t_{cool}$
For the range of production rates estimated in this work (shown in
table~\ref{ratestritium}) and the deduced tritium activities in D0
and D2, the exposure times are between 0.8 and 1.6~years and 4.2 and
8.4~months, respectively. These values roughly agree with the time
lapse between sodium iodide raw material purification starting and
detector shipment, according to the company. As an additional check,
for these exposure times, the ratio of the induced initial
activities of the also long-living cosmogenic isotope $^{22}$Na in
D0 and D2 following equation~\ref{eqac} is $\sim$2, in good
agreement with the measured activities
(159.7$\pm$4.9)~kg$^{-1}$d$^{-1}$~\cite{cebrianjcap} and
(70.2$\pm$3.9)~kg$^{-1}$d$^{-1}$~\cite{anaisepjc}. Since
purification methods cannot remove this isotope, this means that the
raw material was not activated before crystal production. Activity
of $^{22}$Na can be estimated through different distinctive
signatures; in particular, once I and Te cosmogenic isotopes have
decayed, it can be quantified by measuring the beta emissions
absorbed in one detector in coincidence with 1274.5~keV depositions
in other one.
% a further exposition of 30 days for the boat trip from US to Spain

\subsection{Argon and Neon}

There is no experimental information of tritium production in argon
or neon gas. As shown in table~\ref{ratestritiumlit}, there were
estimates of the production rate in argon~\cite{mei,mei2016} but
there was no information for neon. As for other targets, the lowest
estimate for argon comes from TALYS due to the limited energy range
considered. For argon, the excitation functions from the two
different libraries used at low and high energies match reasonably
well (see figure~\ref{efAr}); therefore, the selected cross-sections
seem more reliable than for other targets and the derived production
rate has a lower uncertainty.

For TREX-DM experiment, a detailed background model has been built,
following the full simulation of the main quantified background
sources (thanks to the assessment of the radiopurity of all the
relevant components) and including the background rejection methods
based on the topological information provided by the readout planes
\cite{trexdmepjc}. The model predicts a level of
$\sim$5~events/keV/kg/d in the 2-7~keV region. Tritium emissions are
fully absorbed in the gas producing a signal indistinguishable from
that of a dark matter interaction. If saturation activity were
reached for tritium, according to production rates in
table~\ref{ratestritium}, it would dominate the expected background
model with a contribution larger than 10~events/keV/kg/d. However,
tritium is expected to be suppressed by purification of gas and
minimizing exposure to cosmic rays of the purified gas should avoid
any problematic tritium activation. On the other hand, in TREX-DM,
mixtures of Ar or Ne with 1-2\%iC$_{4}$H$_{10}$ at 10~b are
foreseen; tritium could not only be cosmogenically induced in the
noble gas, but also present in the isobutane. No specific
information about tritium content in isobutane has been found.
Assuming similar concentration as in water\footnote{For natural
surface waters there are about one tritium atom per 10$^{18}$ atoms
of hydrogen, following Ref.~\cite{sourcesoftritium}. The measured
tritium activity in some waters and the limits for drinking water
give indeed higher concentrations.}, this would also give a very
relevant contribution in the RoI of TREX-DM of few tens of
events/keV/kg/d for Ar and Ne.
%In~\cite{mei2016}, the production rate of $^{3}$H at sea level for C$_{2}$H$_{6}$ is estimated from
%GEANT4 simulations as 279.5~kg$^{-1}$ d$^{-1}$; considering the
%saturation activity and if tritium could not be removed, this value
%would correspond to a tritium concentration of 0.9$\times$10$^{-19}$
%atoms per atom of hydrogen, which is roughly one order of magnitude
%lower than the tritium concentration in water.
In any case, obtention of isobutane from underground gas sources
shielded from cosmic rays will avoid a dangerous tritium content.
The first experimental data in TREX-DM, expected for 2018, would be
extremely useful to confirm that tritium production is not a
relevant background source for the experiment. Further experimental
input would help to validate or improve the approach followed in
this work.

\section{Conclusions}
\label{con}

The production of tritium due to the exposure to cosmic rays at sea
level has been studied for materials used in dark matter detectors,
since, as shown by several experimental results, it can become a
very relevant background source due to its very low energy beta
emissions and long half-life. Production rates have been calculated
for germanium, sodium iodide, argon and neon taking into account the
neutron spectrum from~\cite{gordon} and selected excitation
functions.

%cross-sections
In the low energy region below 100~MeV, for all the targets,
production cross-sections for neutrons are larger than for protons,
which confirms the need to use data specific for neutrons in this
energy range. Here, all available neutron data up to 200~MeV have
been taken into consideration. The experimental data for
cross-sections found, even if scarce, validate the TENDL-2013
results considered. For all the targets, cross-sections below
200~MeV increase with energy and therefore contribution from
energies above cannot be neglected. Assuming that for energies of a
few hundreds of MeV yields by neutrons and protons are similar,
cross-sections from HEAD-2009 library can be used; there is in some
cases (germanium and iodine) an important mismatch between
cross-section from the two libraries, which produces an important
uncertainty in the calculation of production rates which has been
evaluated. For light targets for which there is no HEAD-2009 result,
contribution above 200~MeV has been evaluated assuming as a constant
value the available highest energy cross-section.

The tritium production rates estimated in this work, summarized in
table~\ref{ratestritium}, are higher than those obtained in previous
calculations based on TALYS~\cite{mei} because of the limited energy
range of this code, but also higher than results from GEANT4 and
ACTIVIA calculations~\cite{mei2016}. It is worth noting that all
these estimates are based on the same cosmic neutron spectrum. If no
special precautions against tritium production were taken, according
to the estimates presented in this work, tritium reaching saturation
in germanium, NaI, Ar or Ne detectors would imply in the region from
1 to 7~keV a raw background rate of 7.2, 7.9, 13.9, and
21.7~events/keV/kg/day, respectively.

The experimental determination of activated tritium in materials is
difficult as it requires low energy threshold detectors and a very
good knowledge of other background components to try to identify
tritium emissions. The measured production rate in natural germanium
by the EDELWEISS collaboration~\cite{edelweisscos} is compatible
with the estimated rate here; for enriched germanium, the agreement
with the measured rate from the {\sc{Majorana Demonstrator}} data is
worse but reasonable. Concerning NaI, there are hints of the
presence of tritium in the ANAIS detectors~\cite{anaisepjc}; the
calculated production rate is compatible with the required exposure
times of crystals to produce the estimated possible activities. The
acceptable agreement of calculated production rates with available
experimental results can be considered as a validation of the
method, followed to evaluate tritium yields in other targets as
argon or neon and that could be applied also for different
materials.

\section{Acknowledgements}
Professor J.A.~Villar passed away in August, 2017. Deeply in sorrow,
we all thank his dedicated work and kindness. We acknowledge
Yu.~A.~Korovin for providing the HEAD-2009 library and LSC and GIFNA
staff for their support. This work has been financially supported by
the Spanish Ministerio de Econom\'ia y Competitividad and the
European Regional Development Fund (MINECO-FEDER) under grants No.
FPA2014-55986-P and FPA2016-76978-C3-1-P; the Consolider-Ingenio
2010 Programme under grants MULTIDARK CSD2009-00064 17 and CPAN
CSD2007-00042; from the European Commission under the European
Research Council T-REX Starting Grant ref. ERC-2009-StG-240054 of
the IDEAS program of the 7th EU Framework Program; and the Gobierno
de Arag\'on and the European Social Fund (Group in Nuclear and
Astroparticle Physics).
%ANAIS FPA2011-23749
%TREX FPA2011-24058 FPA2013-41085-P
% ARAID Foundation and C.~Cuesta predoctoral
%grant). P.~Villar are supported by the MINECO Subprograma de
%Formación de Personal Investigador.

%% The Appendices part is started with the command \appendix;
%% appendix sections are then done as normal sections
%% \appendix

%% \section{}
%% \label{}

%% References
%%
%% Following citation commands can be used in the body text:
%% Usage of~\cite is as follows:
%%  ~\cite{key}         ==>>  [#]
%%  ~\cite[chap. 2]{key} ==>> [#, chap. 2]
%%

%% References with bibTeX database:
%\bibliographystyle{elsarticle-num}
%\bibliography{<your-bib-database>}

%% Authors are advised to submit their bibtex database files. They are
%% requested to list a bibtex style file in the manuscript if they do
%% not want to use elsarticle-num.bst.

%% References without bibTeX database:

% \begin{thebibliography}{00}
%% \bibitem must have the following form:
%%   \bibitem{key}...
%%
% \bibitem{}
% \end{thebibliography}

\end{document}